# Radio observations of the $\gamma$-ray quasar 0528+134

## Superluminal motion and an extreme scattering event

**M. Pohl**[1,2], **W. Reich**[2], **T.P. Krichbaum**[2], **K. Standke**[2,3], **S. Britzen**[2], **H.P. Reuter**[4], **P. Reich**[2], **R. Schlickeiser**[2], **R.L. Fiedler**[5], **E.B. Waltman**[5], **F.D. Ghigo**[6], **and K.J. Johnston**[7]

[1] Max-Planck-Institut für Extraterrestrische Physik, Postfach 1603, 85740 Garching, Germany
[2] Max-Planck-Institut für Radioastronomie, Auf dem Hügel 69, 53121 Bonn, Germany
[3] Geodätisches Institut der Universität Bonn, Nussallee 17, 53121 Bonn, Germany
[4] Instituto de Radio Astronomia Millimetrica, Avda. Divina Pastora 7, 18012 Granada, Spain
[5] Remote Sensing Division, NRL, Code 7210, Washington, DC 20375-5351
[6] National Radio Astronomy Observatory, P.O. Box 2, Green Bank, WV 24944
[7] US Naval Observatory, 3450 Massachusetts Ave., NW, Washington, DC 20392-5420

[the date of receipt and acceptance should be inserted later]

**Abstract.** We report on multifrequency radio observations made with the Effelsberg 100-m telescope, the IRAM 30-m telescope and the Green Bank Interferometer between 1992 and 1994 of the $\gamma$-ray quasar 0528+134. We present a new VLBI based map of 0518+134 at 22 GHz with sub-mas angular resolution observed in November 1992. At that time the source was in a phase of brightening at all of our observing frequencies above 3 GHz. The increase of brightness may be related to activity in the unresolved core component of the quasar.

The VLBI map at 22 GHz (epoch 1992.85) shows a one-sided core jet structure of $\sim$ 5 mas length. A new component close to the core indicates an apparent transverse velocity of $\beta_{app} \leq 20$ ($H_0$=100 km sec$^{-1}$ Mpc$^{-1}$, $q_0$=0.5). A second component, seen also previously at 8.4 GHz & 22 GHz, shows apparent superluminal motion with $\beta_{app}$=4.4$\pm$1.7. Superluminal motion is expected since strong Doppler boosting in this source is required in view of the $\gamma$-ray luminosity and the variability timescale, which violate the compactness limit.

In summer 1993 we observed a major outburst at all frequencies higher than a few GHz, which peaked a few months after a strong outburst in high energy $\gamma$-rays and showed the canonical time evolution of a cooling and/or expanding electron distribution. Our data indicate that the outburst in the EGRET range originated very close to the central object of the AGN and that a remnant of this outburst moved further outward in the jet until it became optically thin at radio frequencies after a few months.

During the flare in July 1993 we observed with the Effelsberg 100-m telescope an unusually strong decrease of the flux density by about 50% at 4.75 GHz and 10.55 GHz and slightly less at 2.695 GHz. This behaviour is also seen in the monitoring data at 2.25 GHz and 8.3 GHz taken with the Green Bank Interferometer (NRL-GBI). The event lasted less than three days at the higher frequencies and more than two weeks at 2.25 GHz. For the case that this event is related to the intense radio flare some geometrical effects like a small variation of the viewing angle of the quasars jet orientated very close to the line of sight are considered, but found to be an unlikely explanation for the observed behaviour. Alternatively, an extreme scattering event by a small dense plasma cloud in the line of sight is able to match the observed time lag in the lightcurves if we take into account the mas-structure of the source and different spectra of the components on the basis of their brightness in the VLBI maps.

The importance of interstellar scattering is stressed as 0528+134 is seen in the direction of the dark cloud Barnard 30 located at 400 pc distance in the Orion complex and that we note frequent fluctuations of the order of 25% at 2.3 GHz / 2.695 GHz, our lowest monitoring frequencies, while the variations at higher frequencies are more smooth.

It is shown that the appearance of the extreme scattering event in the light curves of 0528+134 depends strongly on its VLBI structure and on the observed spectral appearance in our two frequency VLBI data. Due to these strong constraints our successful modelling provides the best available evidence that unusual variability behaviour of AGN may be caused by extreme scattering events and that localised (r<1 AU) and dense ($n_e$=100-1000 cm$^{-3}$) plasma structures do exist in the interstellar medium of our Galaxy.

*Send offprint requests to*: Martin Pohl

## 1. Introduction

The quasar 0528+134 is one of the strongest AGN in the hard $\gamma$-ray range as observed by EGRET on board of CGRO (Hunter et al. 1993). It is a m = 20.0 red object located in the direction of the dark cloud Barnard 30 at the periphery of the $\lambda$-Orionis complex which is at a distance of about 400 pc (Maddalena et al. 1986). The redshift of 0528+14 has been recently measured to be z = 2.07 by Hunter et al. (1993), which makes the source one of the most luminous objects in the universe.

0528+134 is a known variable radio source and has been regularly monitored for about 20 years (e.g. Aller et al., 1985). After the launch of CGRO in 1991 it was one of the first new detection's by the hard $\gamma$-ray instrument EGRET. The source is located in the galactic anticenter close to the well known $\gamma$-ray sources Geminga and the Crab. Although this area was also deeply observed by COS B, 0528+134 was never noted during that mission lasting from 1975 until 1982. Since the fall of 1991 regular observations with the Effelsberg 100-m telescope have been made in the frequency range 2.695 GHz to 32 GHz. The source has also been monitored by the IRAM 30-m telescope around 90 GHz and less frequently at higher frequencies. Shortly after its detection in the $\gamma$-ray range the source increased substantially in intensity first at 90 GHz and 32 GHz and with some delay also at lower frequencies. Reich et al. (1993) have already presented these observations until the end of 1992. However, the limited time sampling so far did not allow us to establish a clear correlation between $\gamma$-ray and radio bands. We discuss the radio lightcurves up to the beginning of 1994 in section 2.

Many of the $\gamma$-ray bright AGN are identified with superluminal sources. 0528+134 was not known as a superluminal source due to an insufficient number of VLBI maps. VLBI observations at 8.4 GHz and 22 GHz performed in 1990/1991 (Zhang et al. 1994, hereafter referred to as Z94) showed a compact source which was interpreted by two alternative models either as a one or two sided core jet structure. In section 3 we present new VLBI observations at 8.4 GHz and 22 GHz done in 1991/1992, which now allow to discriminate between the different signatures of the VLBI components and to obtain a better understanding of the kinematics of the mas-structure.

Section 4 comprises a general discussion of the results. We found a rather unusual strong effect in the radio light curves in July 1993 when the source flared at 32 GHz and 86 GHz. This event is discussed in some detail in section 4. There we investigate the possibility of an intrinsic effect within the source to explain our observations. Alternatively we consider a scattering effect caused by the galactic interstellar medium in the line of sight to explain the observed radio light curves.

## 2. Multifrequency radio observations

Reich et al. (1993) have already described the multifrequency observing method used at the Effelsberg 100-m telescope to monitor variable sources as detected by EGRET. The observations result in quasisimultaneous flux density measurements at 2.695 GHz, 4.75 GHz, 10.55 GHz and - if weather allows - in addition at 32 GHz. The angular resolution of the telescope is 4.3', 2.4', 1.2' and 0.45' respectively. Orthogonal cross-scans are made centered on the source and a Gaussian fit is applied to the averaged forward and backward scan taken in each direction. Small pointing offsets are corrected for by the analysis software when calculating the peak flux density of the source. A detailed error analysis from the Gaussian fit parameters is made and the scatter in the data obtained from the calibration sources during one observing run is added for the total error of the flux densities quoted. As listed in Table 1 the errors at 2.695 GHz and 4.75 GHz are quite small typical around 0.1 Jy, while they increase at higher frequencies. The observations for one source at all frequencies are typically completed within 30 minutes. 3C286 serves as the main flux density calibrator at all frequencies assuming 10.4 Jy, 7.5 Jy, 4.5 Jy and 2.1 Jy at 2.695 GHz, 4.75 GHz, 10.55 GHz and 32 GHz. In the case of 0528+134 the nearby secondary calibrator 3C138, which is just 4 degrees apart, is always observed before or after measuring 0528+134. This gives an additional check on the accuracy of all high frequency observations of 0528+134, which in general may suffer occasionally by small opaque clouds in the line of sight.

We have also taken data at 2.25 and 8.3 GHz from the NRL-GBI monitoring program to complement our data base. The purpose of this observing program is to study extreme scattering events as described in Fiedler et al. (1994).

The method of observing with the Green Bank Interferometer was already described by Fiedler et al. (1987b) and Waltman et al. (1991). The observing frequencies are 2.25 and 8.3 GHz, and the maximum baseline of the interferometer is 2.4 km. The source was observed on alternate days near the meridian. Calibration was based on four sources but ultimately referred to 3C286. The rms-errors of the derived flux densities are around 2% at 2.25 GHz and roughly 5% at 8.3 GHz.

Steppe et al. (1988) have already described the observing and calibration method at the IRAM 30-m telescope to monitor variable sources. For 0528+134 mainly 86 GHz observations have been made in 1993. Flux density errors are below 8%. Z94 presented closely sampled observations at 22 GHz and 37 GHz made between mid 1991 until 1993 which complement the Effelsberg and IRAM data. The spectra at all wavelengths below 32 GHz are inverted. The lightcurves based on Effelsberg and IRAM data from the end of 1991 until mid the spring in 1994 are shown in Fig.1. In Fig.2a we give a detailed view of the Effelsberg data and compare them with the NRL-GBI data in Fig.2b, which we show as averages over two weeks to improve the readability of the plot. Here the uncertainty in the Green bank data is not due to observational errors, but to day-to-day fluctuations of the source which may sum up to two or three percent. Unfortunately, for a period of two weeks in the first half of July 1993 no data at 2.25 GHz from the Green Bank Interferometer are available due to hardware malfunction.

Fluctuations at 2.695 GHz of about 20% - much larger than the errors in the observed flux densities - are rather frequent during the time of monitoring and these fluctuations are sig-nificantly larger than those at 4.75 GHz, 8.3 GHz, and 10.55 GHz. There is an excellent agreement between the Effelsberg 2.695 GHz and the Green Bank 2.25 GHz data. This is an additional indication that most of the fluctuations seen at this low frequency have to be considered as real variations seen on time scales of months. This is a strong indication that stochastic interstellar scattering is important in the direction of 0528+134, which is not unexpected considering the structure of the interstellar medium in the Orion region.

For a few months in fall 1992, when we made our 22 GHz VLBI observations, the mm-flux densities are around 4.5 Jy, which is lower than the peak values seen in mid 1992 which range between 5.5 Jy and 7 Jy. At the time of the $\gamma-$ray flare in March 1993 the 32 GHz flux was at a high level of about 7 Jy. Unfortunately 86 GHz data are missing during that period. After a short time of slightly lower flux density level at 32 GHz we note in mid 1993 the absolute flux density maximum ranging between 8.5 Jy and 9 Jy at 86 GHz and 32 GHz. At the time of highest flux density at 32 GHz the 86 GHz flux densities declined already and this may be indicative of a time lag. The 10.55 GHz and 8.3 GHz flux densities remained roughly constant at a high level until the end of 1993. The evolution of this source during the outburst follows the canonical behaviour of appearing first at high frequencies, and then with decreasing optical depth being affected at lower and lower frequencies.

A very unusual behaviour, however, is the dramatic drop in intensity at 10.55 GHz and at all of our lower monitoring frequencies at that time in July 1993. In Table 1 we have listed the detailed results of our observations made in 1993. As evident from Table 1 and Fig. 2a there has been a dramatic decrease in flux density by about 50% at 10.55 GHz and 4.75 GHz and slightly less at 2.695 GHz. This happens within four weeks although details are missing by insufficient sampling. Therefore the observed minimum in particular at 10.55 GHz might be even below what we have noted for July 17. The rapid increase of the 10.55 GHz flux density between July 17 and July 18 by about 1 Jy or about 25% of its absolute flux density is remarkable. On July 18 two observations were made revealing a flux density increase by about 0.2 Jy within one hour. At 4.75 GHz and 2.685 GHz the observed flux densities are similar on July 17 and 18. A slightly decreasing flux density at 4.75 GHz within the errors is observed. In view of the observational errors given in Table 1 this decrease is highly significant at more than $10\sigma$ at 4.75 GHz. Our cross checks on the nearby calibrator 3C138 prove that this feature cannot be due to systematic effects. The repeated observations on July 18 and July 23 demonstrate the reliability of our observations. At 10.55 GHz the source recovered within days, at 4.75 GHz within a week, and at 2.695 GHz in less than four weeks.

The same behaviour is seen in the data taken by the Green Bank Interferometer. At 8.3 GHz the flux density was extremely low at the time we observed the depression at 10.55 GHz with the Effelsberg telescope (see also Fig.5). However, since only one observation is strongly influenced the feature does not appear in the two week averages in Fig.2b. At 2.25 GHz the flux is extremely low for a time of roughly two weeks, and the significance of the depression seen in Fig.2b is much higher than $10\sigma$. This is an independent confirmation that the Effelsberg data show a real feature which is only undersampled. The differences in the flux variations at the different frequencies are important constraints for the modelling described below.

The light curves of 0528+134 show that a similar event but with a smaller amplitude may have occurred in June 1992. There is also some evidence for a counterpart in the Green Bank data.

## 3. VLBI Observations

### 3.1. VLBI observations at 22 GHz, Data analysis and Results

0528+134 was observed with a global VLBI array of eleven stations (see Table 2) at 22 GHz from November 7, 21h30 UT, to November 8, 13h30 UT, in 1992. The data were recorded using the Mk III VLBI recording system (Rogers et al., 1983) in mode B with an observing band of 28 MHz centred at 22236.99 MHz. The correlation was done at the Max-Planck-Institut für Radioastronomie at Bonn. Fringe-fitting and imaging was done in the usual way using standard VLBI-software (Alef 1989, Pearson, 1991, Shepherd et al., 1994). To correct for atmospheric attenuation, the amplitude calibration of the visibilities was improved applying opacity corrections to each station (for details of the method see Krichbaum et al., 1993). In Table 2 the participating stations and their antenna performance are summarised.

In Figure 3 we show the CLEAN-map of 0528+134 obtained at 22 GHz. The source structure can be described as a one-sided bent core-jet structure of $\sim$ 5 mas length. The parameters of the individual structure components were determined from fits of elliptical Gaussian components to the data. The best results were obtained with the 6 component model given in Table 3a. In this model distinct jet components were found at separations of $r = 0.2, 0.6, 0.9, 3.1,$ and $4.9$ mas relative to the brightest component. Due to their faintness the two outer components at $r = 3.1$ mas and $r = 4.9$ mas are less well defined than the 3 central components and may reflect the possibility that the jet emission at separations of $r > 3$ mas is resolved by the interferometer beam and/or just gets too faint (the map contains about 90% of the total flux density). During the model fitting some uncertainty remained in the determination of the exact location of the emission between $r = 0.6$ mas and $r = 1.0$ mas. A slightly worse model fit with only 5 components, yields an extended component C1/C2 at $r = 0.8$ mas, which concatenates C1 and C2 into a single component. In Table 3a the parameters of this concatenated component C1/C2 are summarised. The non-uniqueness of the models may reflect a relatively sparse uv-coverage on intermediate-length baselines. Since the data obtained at 8.4 GHz (see below) also showed distinct emission near $r = 0.5$ and $r = 1.0$ mas, we used the more complex 6 component model for further discussions.

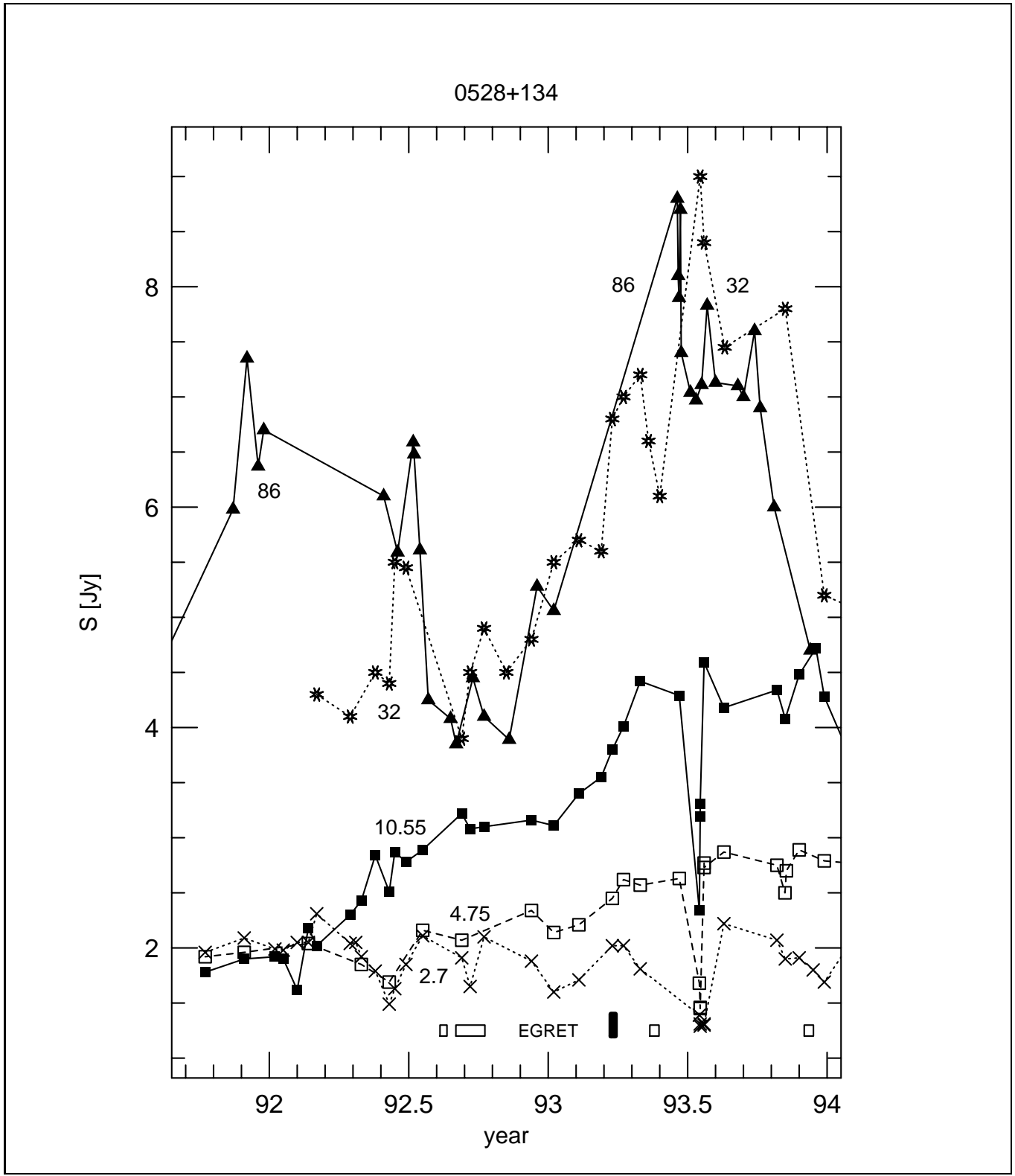

**Fig. 1.** Lightcurves of 0528+134 from the Effelsberg 100-m and the IRAM 30-m telescopes. The numbers are the corresponding observing frequencies in GHz. The errors of the individual data are given in Table 1. The source activity in the EGRET range is indicated by open rectangles for normal level and filled rectangles for a high $\gamma$−ray flux.

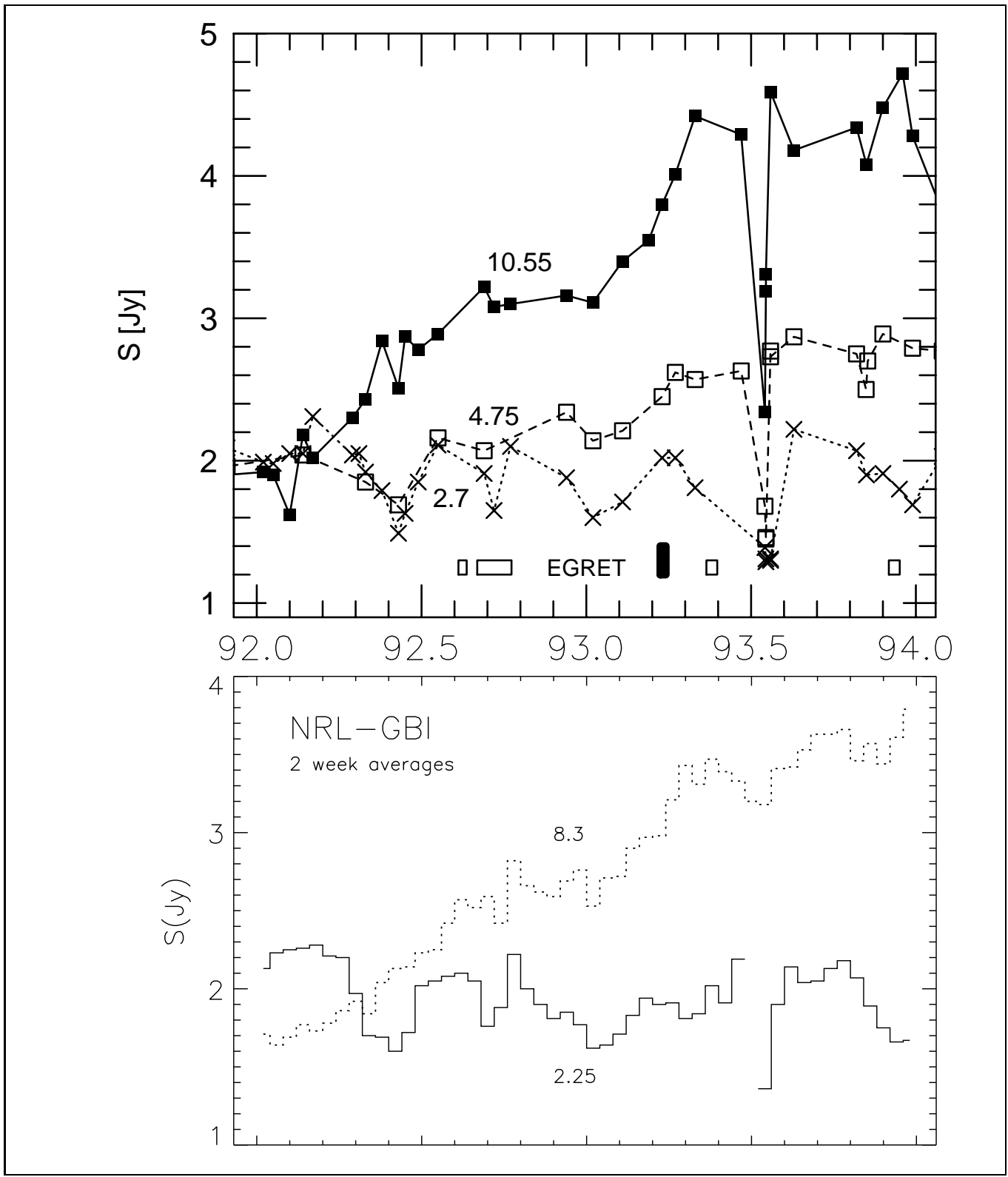

**Fig. 2a.** (On top) Details of the lightcurves of 0528+134 at 2.695 GHz, 4.75 GHz and 10.55 GHz. Again the $\gamma$−ray state is indicated.

**Fig. 2b.** (Bottom) The lightcurves of 0528+134 at 2.25 GHz and 8.3 GHz as averages over periods of two weeks. For the two weeks in July 1993 no data at 2.25 GHz are available due to hardware malfunction. The dip at 8.3 GHz in July 93 is averaged out since only one of six observations is influenced.

**Table 1.** Flux densities of 0528+134 in 1993 as observed with the Effelsberg 100-m telescope and the IRAM 30-m telescope. Frequencies are in GHz and flux densities in Jy.

| Date | S (2.69) | dS (2.69) | S (4.75) | dS (4.75) | S (10.55) | dS (10.55) | S (32) | dS (32) | S (86) | dS (86) |
|---|---|---|---|---|---|---|---|---|---|---|
| 06.01. |  |  |  |  |  |  |  |  | 5.1 | 0.2 |
| 08.01. | 1.60 | 0.06 | 2.14 | 0.07 | 3.11 | 0.09 | 5.5 | 0.6 |  |  |
| 09.02. | 1.71 | 0.07 | 2.21 | 0.12 | 3.40 | 0.24 | 5.7 | 0.5 |  |  |
| 09.03. |  |  |  |  | 3.60 | 0.23 | 5.6 | 0.3 |  |  |
| 11.03. |  |  |  |  | 3.50 | 0.21 |  |  |  |  |
| 24.03. | 2.02 | 0.08 | 2.45 | 0.11 | 3.80 | 0.23 | 6.8 | 0.4 |  |  |
| 09.04. | 2.02 | 0.10 | 2.62 | 0.14 | 4.01 | 0.20 | 7.0 | 2.0. |  |  |
| 30.04. |  |  |  |  |  |  | 7.2 | 0.5 |  |  |
| 01.05. | 1.81 | 0.07 | 2.57 | 0.14 | 4.42 | 0.30 |  |  |  |  |
| 10.05. |  |  |  |  |  |  | 6.6 | 0.7 |  |  |
| 25.05. |  |  |  |  |  |  | 6.1 | 0.5 |  |  |
| 18.06. |  |  |  |  |  |  |  |  | 8.8 | 1.0 |
| 19.06. |  |  |  |  |  |  |  |  | 8.1 | 0.9 |
| 20.06. |  |  | 2.63 | 0.08 | 4.29 | 0.21 |  |  | 7.9 | 0.9 |
| 22.06. |  |  |  |  |  |  |  |  | 8.7 | 1.0 |
| 23.06. |  |  |  |  |  |  |  |  | 7.4 | 0.8 |
| 05.07. |  |  |  |  |  |  |  |  | 7.0 | 0.4 |
| 13.07. |  |  |  |  |  |  |  |  | 7.0 | 0.4 |
| 17.07. | 1.39 | 0.04 | 1.68 | 0.08 | 2.34 | 0.16 |  |  |  |  |
| 18.07. | 1.31 | 0.04 | 1.46 | 0.06 | 3.19 | 0.15 | 9.0 | 0.6. |  |  |
| " +1h | 1.29 | 0.03 | 1.45 | 0.06 | 3.31 | 0.17 |  |  |  |  |
| 19.07. |  |  |  |  |  |  |  |  | 7.1 | 0.2 |
| 23.07. | 1.30 | 0.04 | 2.77 | 0.11 | 4.59 | 0.26 | 8.4 | 0.8 |  |  |
| " +1h | 1.31 | 0.04 | 2.73 | 0.11 | 4.59 | 0.25 |  |  |  |  |
| 26.07. |  |  |  |  |  |  |  |  | 7.8 | 0.5 |
| 08.08. |  |  |  |  |  |  |  |  | 7.1 | 0.2 |
| 18.08. | 2.22 | 0.11 | 2.87 | 0.12 | 4.18 | 0.27 |  |  |  |  |
| 19.08. |  |  |  |  |  |  | 7.45 | 0.35 |  |  |
| 04.09. |  |  |  |  |  |  |  |  | 7.1 | 0.3 |
| 12.09. |  |  |  |  |  |  |  |  | 7.0 | 0.2 |
| 28.09. |  |  |  |  |  |  |  |  | 7.6 | 0.3 |
| 04.10. |  |  |  |  |  |  |  |  | 6.9 | 0.2 |
| 21.10. |  |  |  |  |  |  |  |  | 6.0 | 0.2 |
| 26.10. | 2.07 | 0.08 | 2.75 | 0.12 | 4.34 | 0.30 |  |  |  |  |
| 06.11. |  |  | 2.50 | 0.11 | 4.05 | 0.27 | 7.8 | 1.0 |  |  |
| 07.11. |  |  | 2.70 | 0.11 | 4.08 | 0.36 |  |  |  |  |
| 08.11. |  |  | 2.71 | 0.11 | 4.10 | 0.26 |  |  |  |  |
| 25.11. | 1.91 | 0.07 | 2.89 | 0.17 | 4.48 | 0.22 |  |  |  |  |
| 10.12. |  |  |  |  |  |  |  |  | 4.7 | 0.3 |
| 12.12. | 1.80 | 0.10 |  |  |  |  |  |  |  |  |
| 16.12. |  |  |  |  | 4.72 | 0.43 |  |  |  |  |
| 27.12. | 1.69 | 0.06 | 2.79 | 0.11 | 4.28 | 0.31 | 5.2 | 0.6 |  |  |

**Table 2.** Station characteristics at 22 GHz. Column 1-6 give the station name and location, the diameter of the antenna, the typical system-temperature during the observations, the peak sensitivity of the antenna gain, and the mean atmospheric zenith-opacity during the experiment, respectively.

| Station | Location | Diam. [m] | $T_{sys}$ [K] | g [K/Jy] | $\tau$ |
|---|---|---|---|---|---|
| Effelsberg | Germany | 100 | 110 | 0.791 | 0.07 |
| Medicina | Italy | 32 | 150 | 0.110 | 0.09 |
| Onsala | Sweden | 20 | 320 | 0.053 | 0.04 |
| Green Bank | West Virginia | 43 | 70 | 0.141 | 0.06 |
| Hancock | New Hampshire | 25 | 90 | 0.087 | 0.05 |
| North Liberty | Iowa | 25 | 100 | 0.108 | 0.09 |
| Fort Davis | Texas | 25 | 60 | 0.095 | 0.06 |
| Los Alamos | New Mexico | 25 | 90 | 0.096 | 0.06 |
| Pie Town | New Mexico | 25 | 100 | 0.134 | 0.07 |
| Kitt Peak | Arizona | 25 | 95 | 0.113 | 0.06 |
| Owens Valley | California | 25 | 150 | 0.108 | 0.07 |

*3.2. VLBI Observations at 8 GHz, Data analysis and Results*

Within the IRIS-S (*International Radio Interferometric Surveying*) and EUROPE geodetic VLBI observing campaigns performed monthly (IRIS-S) and quarterly (EUROPE), the source 0528+134 is observed regularly in the S- and X-band (2.3 GHz & 8.4 GHz). Geodetic VLBI observations therefore provide an important *astronomical* data base to study the mas-structure of compact extragalactic radio sources and its kinematics on timescales of a few months (Britzen et al., 1993). From these observing campaigns we selected and analysed two data sets in the X-band (epochs: February 25, 1991 (1991.23) & March 25, 1992 (1992.15)), which were relatively close in time to our observations at 22 GHz and previous VLBI observations performed in 1990/1991 by Z94. In these experiments 0528+134 was observed in snapshot mode for 12 hrs using the VLBI antennas at Mojave (USA), Richmond (USA), Westford (USA), Wettzell (Germany) and Harthebesthoek (South Africa). Mainly due to the participation of the antenna at South Africa, the observing beam was nearly circular with a size of typically $\sim 0.5$ mas. The data were correlated and imaged with the same standard software as described above. The amplitude calibration was done in the standard way, using regular system-temperature measurements and antenna gain curves provided by the stations. The 'a priori' amplitude calibration based on data provided by each observatory was further improved by imaging of structure calibrators (sources of simple and known mas-structure, e.g. 1803+784, 4C39.25, see Britzen et al., 1994) in parallel to 0528+134, using methods described in Schalinski et al., 1986. The maps and model fits obtained from the two data sets for 0528+134 clearly show a one-sided core-jet structure of more than 5 mas length, bending from $P.A. \simeq 60°$ near the core to $P.A. \simeq 5° - 20°$ at $r \geq 5$ mas. The mas-structure at 8 GHz can be best described by 5 distinct features C0 - C4, which are located (relative to the brightest component C0) at separations of $r = 0.34 \pm 0.18$ mas (1991.23) and $r = 0.57 \pm 0.18$ mas (1992.15), respectively, $r \simeq 1.1$ mas, $r \simeq 3.2$ mas, and $r \simeq 4.9$ mas. In Table 3b & 3c we summarize the model fit parameters of the VLBI components at both epochs.

*3.3. Discussion of the source structure*

Inspection of maps and model fits at 8 GHz and 22 GHz reveals a core-dominated core-jet structure, with embedded distinct features and an overall pronounced and continuous bending of about $60° - 70°$: the elliptically elongated component C0 and the innermost jet component (at $r = 0.2$ mas) are oriented along $P.A. = 80°$, the components near $r = 0.6$ mas are at $P.A. \simeq 60°$, the components near $r = 1.0$ mas are at $P.A. \simeq 30°$, and the outer components are located at $P.A. = 10° - 25°$.

3.3.1. Identification of the VLBI core:

The interpretation of the morphology and kinematics of the mas-structure of AGN critically depends on the existence of distinct jet components, registered at the different observing times and frequencies. Little was known up to now on the mas-structure of 0528+134. A first map published by Charlot (1990) revealed a first but only crude estimate of the basic source structure. More detailed maps obtained in 1990/1991 at 22 GHz and 8 GHz by Z94 yielded two alternative component identifications schemes, which basically differed in the interpretation of the brightest component as being or not being the VLBI-core of 0528+134. With the new data, it is now possible to discriminate between these two schemes.
Based on our results summarised in Table 3 and the maps and model fits obtained by Z94 (1990.85: 22 GHz map, 1991.16: 8 GHz map) we suggest the component identification indicated by label 'C' in column 1 of Table 3, in which C0 is the VLBI core, which we assume to be stationary. This identification of C0 as VLBI core is supported by its pronounced flux density variability, its inverted spectrum, the one-sidedness of the mas-structure, and the fact that two components (N & C1) seem to move away from C0. To include the VLBI components given by Z94 in the discussion, we redetermined their positions relative to the concatenated components K2K3 and X2X3 (see Table 1 in Z94), which we identified as the core. Preliminary results obtained from VLBI observations at 43 GHz and 86 GHz support this identification scheme (Standke 1994).

3.3.2. Motion of C1:

With the assumption of C0 being the stationary VLBI core a *simple* cross-identification of the structure components seen in the maps/model fits at 8 & 22 GHz is suggested. In this identification scheme the components C2 (at $r \simeq 1.0$ mas), C3 (at $r \simeq 3.2$ mas) and C4 (at $r \simeq 4.9$ mas) remained stationary relative to C0 within the observing interval 1991.16 - 1992.85. For component C1, however, we find evidence for an increase of its core separation with an apparent angular separation rate of $\mu = 0.11 \pm 0.043$ mas/yr (see Figure 4), corresponding to superluminal motion of $\beta_{app} = 4.4 \pm 1.7$ (for $z = 2.07$ and $H_0 = 100$ km s$^{-1}$ Mpc$^{-1}$, $q_0 = 0.5$). Subluminal motion is rejected with more than $2\sigma$ (4.2% chance probability). The rejection is even stronger for a smaller Hubble constant $H_0 < 100$ km s$^{-1}$ Mpc$^{-1}$. A linear fit for the angular separation rate has a significance level of only 57% in the $\chi^2$-statistics. This is mainly caused by the small error bars for the earlier data by Zhang et al. (Z94), which partly are the result of the small FWHM of the component C1 at this time. The error bars provided for our VLBI observations were calculated including not only the the formal Gaussian errors, but also uncertainties derived from the internal scatter between slightly different calibrated data sets and the resulting slightly different models of the source structure, and thus may be rather too conservative.

**Table 3a.** Results of the Gaussian component model fits at 22.2 GHz, 1992.85. Here and in the tables 3b & 3c column 1 gives the component identification, column 2 the flux density, column 3 and 4 the component position in polar co-ordinates (distance, position angle), column 5 the geometric mean size of the major axis and minor axis of the Gaussian component. Numbers in parentheses are formal errors assuming Gaussian error propagation. The position angles are counted from north through east.

| Id. | $S$ [Jy] | $r$ [mas] | $P.A.$ [°] | $FWHM$ [mas] |
|---|---|---|---|---|
| C0   | 3.00 (0.05)    | 0          | 0       | 0.08 (0.02) |
| N    | 0.32 (0.03)    | 0.23 (0.03)| 83 (10) | 0.21 (0.10) |
| C1   | 0.16 (0.05)    | 0.63 (0.20)| 64 (20) | 0.72 (0.20) |
| C2   | 0.10 (0.05)    | 0.92 (0.20)| 30 (10) | 0.61 (0.20) |
| C3   | 0.017 (0.008)  | 3.14 (0.35)| 27 (5)  | 0.20 (0.13) |
| C4   | 0.016 (0.011)  | 4.9 (0.9)  | 23 (10) | 0.76 (0.90) |
| C1/2 | 0.22 (0.05)    | 0.81 (0.09)| 44 (5)  | 1.01 (0.19) |

**Table 3b.** Results of Gaussian component model fits at 8.4 GHz, 1991.23.

| Id. | $S$ [Jy] | $r$ [mas] | $P.A.$ [°] | $FWHM$ [mas] |
|---|---|---|---|---|
| C0 | 0.79 (0.20) | 0          | 0       | 0.44 (0.10) |
| C1 | 0.24 (0.09) | 0.34 (0.18)| 58 (18) | 0.20 (0.10) |
| C2 | 0.33 (0.15) | 1.13 (0.23)| 24 (7)  | 0.60 (0.20) |
| C3 | 0.10 (0.05) | 3.32 (0.26)| 24 (5)  | 0.43 (0.30) |
| C4 | 0.09 (0.06) | 4.8 (0.5)  | 6 (10)  | 0.54 (0.25) |

**Table 3c.** Results of Gaussian component model fits at 8.4 GHz, 1992.15.

| Id. | $S$ [Jy] | $r$ [mas] | $P.A.$ [°] | $FWHM$ [mas] |
|---|---|---|---|---|
| C0 | 1.24 (0.07) | 0          | 0       | 0.37 (0.10) |
| C1 | 0.21 (0.12) | 0.57 (0.18)| 57 (5)  | 0.10 (0.05) |
| C2 | 0.10 (0.05) | 1.09 (0.10)| 25 (10) | 0.63 (0.20) |
| C3 | 0.12 (0.07) | 3.08 (0.37)| 22 (6)  | 0.31 (0.15) |
| C4 | 0.11 (0.07) | 5.0 (0.4)  | 12 (10) | 0.29 (0.15) |

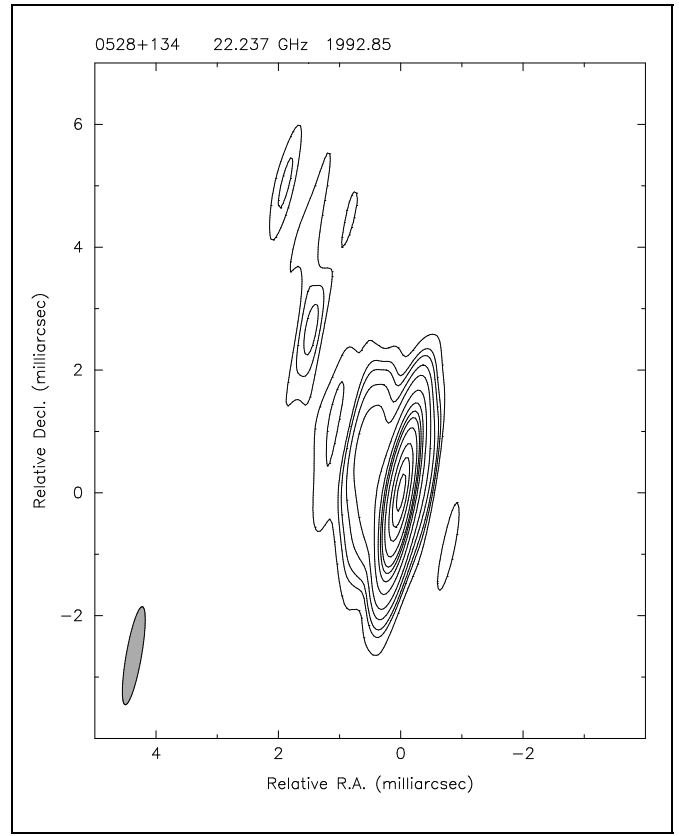

**Fig. 3.** Clean-Map of 0528+135 at 22 GHz restored with an elliptical beam of size 1.62 x 0.25 mas, $pa = -10°$. Contour levels are -0.1, 0.1, 0.3, 0.5, 1, 2, 5, 10, 15, 20, 30, 50, 70, and 90 % of the peak flux density of 2.81 Jy/beam. The brightest component C0 corresponds to the core. In its eastern wing the three components N, C1, and C2 are hidden, whereas the weaker outer components C3 and C4 can be clearly seen in the north-west of the core. The structure east of the core is at the noise level. A simple sketch of the inner structure will be shown in Fig.6.

Another source of uncertainty arises when using data at different frequencies, namely a possible frequency dependence of the core separation. For 0528+134 this effect appears to play a minor rôle as the 8.4 GHz data and the 22 GHz and 43 GHz data in Fig.4 follow – within the statistics – the same regression line. However, it may be wortwhile to check the stability of our results against this systematic uncertainty by enlarging the error bars to at least $\delta r = 0.1$ mas. Doing so for the two earlier data points does not change much of the result, but improves the $\chi^2_{min}$ to 95% significance level. However, then subluminal motion is rejected with only $1.8\sigma$ (7.3% chance probability). A preliminary analysis of 8.4 GHz VLBI observations at epoch 1993.53 yields an indication that the component C1 is a blend of two individual components (Britzen et al., in prep.). These new observations have been done in the same manner as the earlier 8.4 GHz observations except that the stations Santiago in Chile and Fortaleza in Brazil replaced the stations at Mojave and Richmond.

It is noteworthy that the position angle of C1 relative to the core C0 changed smoothly from $P.A. \simeq 89°$ in 1990.85 (Z94) to $P.A. \simeq 50° - 60°$ in 1992 (see Table 3). Since the jet is oriented along $P.A. \simeq 10° - 20°$ at larger core separations, it is tempting to assume that C1 follows a bent path. With regard to the observed bent jet structure (See Fig. 3) such a behaviour would not be completely unexpected and is also seen with mm-VLBI in an increasing number of other blazars (e.g. Krichbaum et al., 1994, and references therein).

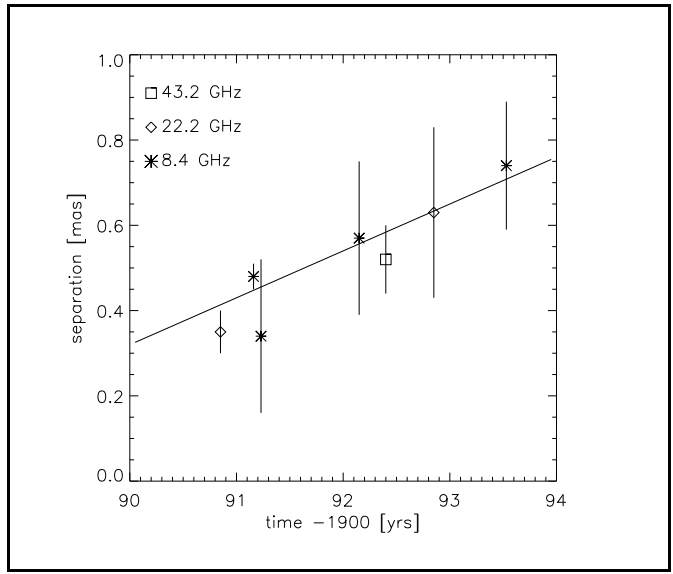

**Fig. 4.** Core separation of component C1 relative to C0. Data points at epochs 1990.85 (22 GHz) and 1991.16 (8 GHz) are from Zhang et al., 1994. The data point at 43 GHz (epoch 1992.4) is taken from Krichbaum et al. (1995). We have further added a data point from a preliminary analysis of 8.4 GHz observations at epoch 1993.53 (Britzen et al., in preparation). The line represents a $\chi^2$-fit to the data with parametrization $r(t) = (0.11 \pm 0.043) \cdot (t - 1990) + (0.32 \pm 0.05)$.

### 3.3.3. Ejection of a new jet component ?

The 22 GHz data revealed the existence of a new component at $r = 0.23 \pm 0.03$ mas, east of the core C0 ($P.A. = 83 \pm 10°$), not seen previously at 8 GHz or 22 GHz, but close to the 1990-position of its precursor C1. We have labelled this component 'N' (see Table 3a). This component is also seen in a 43 GHz map from 1992.40 (Krichbaum et al., 1995) and at 86 GHz from 1993.3 (Standke, 1994). It is therefore likely that a new jet component was ejected prior to 1992.4 and is now moving outwards. Disregarding possible small frequency shifts due to opacity effects between 22, 43, and 86 GHz, we tentatively determined the angular separation rate of 'N' to be about $\leq 0.5$ mas/yr ($\beta_{app} \leq 20$). Future observations should clarify, whether component 'N' was ejected during the high-frequency radio outburst in 1992.1.

### 3.3.4. Spectral indices of the jet components:

With the available data it is possible to determine the spectral shape of the jet components. In Table 4 we summarise the results for the spectral indices $\alpha_{8/22\,GHz}$ and $\alpha_{22/43\,GHz}$. The errors for the spectral indices were calculated formally, using the errors of the corresponding flux densities given in Table 3 and assuming Gaussian error propagation. Between 1990/1991 and 1992 the spectrum of the core component C0 inverted considerably, reflecting that the flux density of the core increased - during an ongoing outburst - first at the higher frequency. With spectral slopes of $\alpha_{8/22\,GHz} = -0.90 \pm 0.10$ and $\alpha_{22/43\,GHz} = 0.30 \pm 0.10$ a spectral turnover near $\nu_m = 22$ GHz could be expected for C0. However, the variability during 1992 in our 32 GHz data and the 22 GHz and 37 GHz data in Z94 imply that this determination of a turnover frequency should not be taken too seriously.

The apparent spectral steepening of the component spectra with increasing core-separation (see Table 4) has to be regarded with some care. Whereas the spectral slopes of the inner jet components N, C1, and C2 are smaller than 0.7, the spectral indices of the components C3 and C4 lie in the range of $1 - 2$. At 22 GHz, however, the outer components C3 & C4 are most probably resolved by the interferometer beam, leading to underestimates of their flux density and therefore to upper limits of the corresponding spectral index $\alpha_{8/22}$. Within still relatively large uncertainties the spectral slopes for the jet components N-C4 therefore are consistent with typical optically thin synchrotron spectra (as observed in comparable objects), and gives us confidence that our interpretation and cross-identification of the features seen at 8 GHz, 22 GHz, and 43 GHz is not implausible.

## 4. Discussion

### 4.1. Luminosity limits and superluminal motion

An interesting aspect of the analysis of AGN $\gamma$-ray data is to look for luminosity limits above or below which isotropic emission from the source is not possible. A violation of these limits can then be taken as argument in favor of an origin of the emission in a relativistic jet pointing nearly towards us. In such a scenario superluminal motion of VLBI components is expected.

Under the synchrotron-self-Compton model Zhang et al. (Z94) have tried to determine the physical conditions of the emission region on the basis of ROSAT data taken in March 1991. If the angular size of the emission region is known (from VLBI) and one is able to get the turnover frequency between optically thick and optically thin synchrotron emission from fits of the radio spectrum, then the self-Compton X-ray flux can be calculated. Z94 found that the predicted X-ray flux is more than three orders of magnitude higher than the flux observed by ROSAT, and conclude that Doppler boosting is required in 0528+134. These authors used a turnover frequency of 7 GHz and a VLBI size for the core of 0.17 mas, which they obtained from 22 GHz VLBI observations taken half a year earlier. Both parameters are very crucial since they enter the prediction by the sixth and eighth power respectively. We have reinvestigated this case and find that at the time of the ROSAT observations the VLBI size at 8.4 GHz was 0.4 mas, both in our observations and in observations by Zhang et al. (Z94), which they did not use for their analysis. This size should be taken as an upper limit, since the core is not resolved; however, it is the only information we have from that time.

**Table 4.** Spectral indices of the components. Column 1: component identification, column 2: observing epochs of frequency1/frequency2 used for $\alpha_{frequency1/frequency2}$, column 3: core separation of the component at epoch 1992.85, column 4 & 5: spectral indices $\alpha_{8/22\,GHz}$ and $\alpha_{22/43\,GHz}$, using $S \propto \nu^{-\alpha}$.

| Id. | interval | r(1992.85) [mas] | $\alpha_{8/22\,GHz}$ | $\alpha_{22/43\,GHz}$ | comment |
|---|---|---|---|---|---|
| C0 | 1990.85/1991.16 | 0 | $0.30 \pm 0.10$ | | from Z94 |
| C0 | 1992.15/1992.85 | 0 | $-0.90 \pm 0.10$ | | |
| C0 | 1992.85/1992.40 | 0 | | $0.30 \pm 0.10$ | using 43 GHz data |
| N | 1992.40/1992.85 | 0.2 | | $0.70 \pm 0.70$ | using 43 GHz data |
| C1 | 1992.15/1992.85 | 0.6 | $0.30 \pm 0.70$ | | |
| C2 | 1992.15/1992.85 | 0.9 | $0.10 \pm 0.10$ | | |
| C3 | 1992.15/1992.85 | 3.1 | $1.6 \pm 0.5$ | | upper limit, see text |
| C4 | 1992.15/1992.85 | 4.9 | $1.9 \pm 0.9$ | | upper limit, see text |

Furthermore, at the time of ROSAT observations the radio flux from 0528+134 was around 2.1 Jy at 2.25 GHz, 1.7 Jy at 8.3 GHz (our NRL-GBI data) and 2.4 Jy at 90 GHz (Reich et al. 1993). Other at least nearly contemporaneous data are not available. Therefore, the turnover frequency in the radio spectrum can not be easily determined. It is, however, surely not 7 GHz. One could argue for a optically thick subcomponent in the core. But also a turnover frequency of 80 GHz with a size of 0.1 mas give a prediction which matches the ROSAT flux. Thus Doppler boosting is not required in view of the low luminosity in the ROSAT range. The assertion by Z94 is no longer tenable.

Recently, McNaron-Brown et al. (1995) argued for beaming of the MeV emission from 0528+134 on the basis of OSSE data. The luminosity and the variability timescale at these energies implied a violation of the compactness limit, i.e. the optical depth against photon-photon pair production was much larger than unity.

Due to the strong reduction of the Compton cross section in the Klein-Nishina limit the Eddington limit in terms of the luminosity and the variability timescale (Elliot and Shapiro 1974) do not force us to the conclusion that the GeV-emission of AGNs is strongly Doppler-boosted. Also the pair production cross section is reduced at photon energies in the EGRET range and hence the corresponding optical depth is less than unity for the interaction between high energy $\gamma$-rays (see also Dermer and Gehrels 1995). However, the compactness limit was violated by 0528+134 in the medium energy $\gamma$-ray range (1-30 MeV), thereby indicating a requirement for Doppler-boosting. A detailed discussion of luminosity limits in the relativistic range, i.e. for $\gamma$-ray energies exceeding the rest mass of electrons, can be found in Appendix B.

Our finding of superluminal motion nicely complements to the requirements for relativistic motion due to the luminosity limits. Therefore it provides strong evidence that blazar radiation is produced by a relativistically outflowing jet which beams emission towards favourably oriented observers.

### 4.2. The lightcurves

In March 1993 0528+134 exhibited an extremely strong $\gamma$−ray outburst ($I(E > 100\,\mathrm{MeV}) \simeq 3 \cdot 10^{-6}\,\mathrm{cm}^{-2}\mathrm{sec}^{-1}$) lasting for a couple of days (Michelson et al. 1994). In May 1993 the $\gamma$-ray flux was down to a normal level similar to the flux observed in fall 1992. We noticed no simultaneous radio flare at that time except for a secondary maximum at 32 GHz. On the basis of monitoring data at 22 Ghz and 37 GHz Z94 concluded that the $\gamma$-ray flare occured just prior to a sharp peak at radio frequencies. With our data we can say that this maximum in the radio flux density was a secondary one and that it occured between the two $\gamma$-ray observations, the first showing an outburst and the second showing low-level emission. In fact the 32 GHz flux density is the same within the errors at the exact time of EGRET observations (see Table 1). The 22 GHz and 37 GHz data by Z94 end roughly at the time of the main radio flare when on average the flux density is higher than for the secondary maximum in April except for one data point at each frequency which may be either flickering or ill-determined.

It may well be that a simultaneous flare occured at optical or NIR wavelengths, but we prefer to draw a relation between the extremely bright γ-ray flare and the main radio flare which became visible a few months later during the summer of 1993 and not between the γ-ray flare and the small secondary maximum in the radio lightcurve.

We have speculated that the new VLBI component N could have been expelled from the core during the outburst at high radio frequencies in the beginning of 1992. Since in summer 1993 the intense radio flare at all frequencies above a few GHz appears to be much stronger than the outburst in the beginning of 1992 it may well be that again a new VLBI component was born, which could be checked by new VLBI observations.

We found a rather unusual strong depression in the radio light curves in July 1993 when the source flared at 32 GHz and 86 GHz. Based on the data presented above we discuss two cases as the most likely explanation for the this strong variability of 0528+134. Firstly we consider a possible origin within the source itself. A different approach suggested by the permanent fluctuations at 2.25 GHz and 2.695 GHz and the dark cloud Barnard 30 along the line of sight is the effect of an extreme scattering event by the interstellar medium.

### 4.3. Intrinsic variations

Obviously, any modelling of the July 1993 event suffers from missing data coverage in the first half of the month. We can therefore not exclude the possibility that the event is caused by intrinsic properties of the source itself. The strong γ-ray emission and variation on very short time scale as well as the available VLBI observations indicate that a relativistic jet is pointed towards us with a very small deviation from the line of sight. In such a geometry even very small bending effects may cause large amplitude fluctuations as has been studied for the case of an anisotropic jet with non-spherical emission regions by Reynolds (1982) and Reynolds and Ellison (1992). Such effects may even lead to intraday variability in the optical synchrotron radiation and in the radio regime. However, since this process could influence only one VLBI component at a time, it appears difficult to reproduce the observed behaviour by changing the viewing angle towards the flaring core component C0. Judged by their intensity in the 22 GHz VLBI map, all components except the core do not provide enough flux at 10.55 GHz to reduce the total radio emission by 2 Jy if they are switched off. Different components should not know what the others do and thus intrinsic variations have to be related to the core component. Due to its optical thickness it is questionable whether the core also provides enough 2.25 GHz and 2.695 GHz flux to explain the observed behaviour. However, at 32 GHz and 86 GHz no variation is observed within an uncertainty of around 10% which can hardly be accounted for in emission models of an anisotropic jet. Additionally, component C0 would have to return to its original path after the event, which is an unlikely behaviour for a bent jet.

One could also think about an optically thick synchrotron blob in the jet passing in front of the core and thereby obscuring part of its emission. The temporal behaviour of the event would then have something to do with frequency dependent substructure in the core. However, a transverse motion by 0.1 mas in about a week implies velocities of around 200 times the speed of light which is impossible. Please note that here the effect of apparent superluminality does not occur.

A third possibility was discussed by Marscher (1979) on the basis of an unpublished suggestion by McAdam. The basic idea is that a layer of H$I$ in front of the source gets ionized by UV photons from an optical outburst and then provides an optically thick screen due to free-free absorption. However, covering the core and the inner components with such a screen requires finetuning to the $10^{-4}$ level, which is the duration of the event divided by the light travel time for the system. Furthermore, the required emission measure EM$\simeq 10^8$ pc cm$^{-6}$ and an extremely dense layer is required to get the recombination time scale to the order of a day.

To summarise, we regard the possibility of an intrinsic cause for the observed variability as unlikely.

### 4.4. Extreme scattering in Barnard 30

In the last decade several studies have focused attention on refractive and diffractive effects of electron density fluctuations in the interstellar medium upon radio observations of pulsars and compact extragalactic radio sources. Both low-frequency intensity variations and time-of-arrival fluctuations in case of pulsars have been successfully modelled as scintillation due to interstellar turbulence (Sieber 1982; Rickett 1986; Cordes, Pidwerbetsky and Lovelace 1986; Romani, Narayan and Blandford 1986; Goodman et al. 1987; Rickett et al. 1994). While these effects are stochastic in nature, the so-called extreme scattering events have provided evidence for the existence of discrete large-scale inhomogenuities in the ionised interstellar medium (Fiedler et al. 1987a; Cognard et al. 1993).

We have already noted that the variability of about 25% at 2.25 GHz and 2.695 GHz on the timescale of a few months gives some evidence for stochastic interstellar scattering (RISS). It is therefore not unlikely that from time to time extreme scattering events also occur.

The mathematical details of our model can be found in Appendix A. The physics involved here is basically the dispersion of electromagnetic radiation in a plasma. The phase velocity of radio waves in a plasma is no longer equal to $c$, but slightly larger. The scatterer thus imposes a phase shift on the traversing waves similar to a diverging lens. The resultant intensity profile can then be calculated following Huygens' principle, that each element of the distorted wavefront is the center of a secondary disturbance which gives rise to spherical wavelets. The position of the wavefront at any later time is determined as the envelope of all such wavelets.

The model curve in Fig. 5 is calculated for a cloud of ionised material with a soft density profile, such that the line-of-sight integral of an electromagnetic wave passing at distance $\rho$ from the cloud centre follows

$$\int_{-\infty}^{\infty} dl \, n_e = \frac{n_e \, r_0^2}{r_0 + \rho}$$

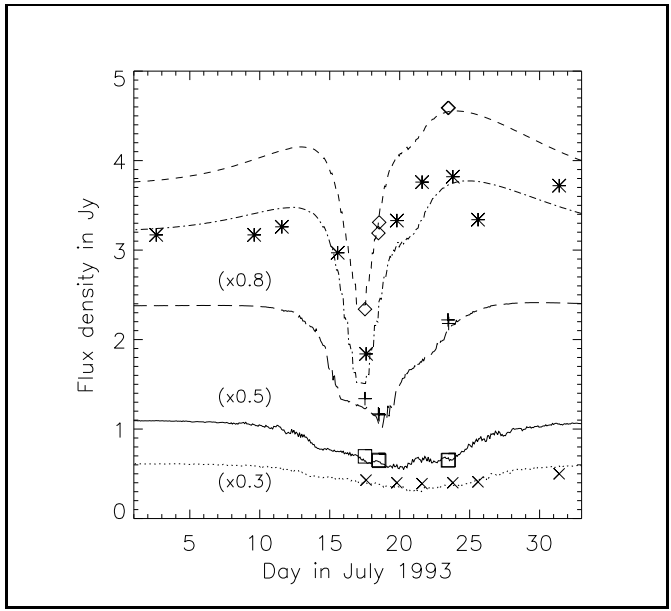

**Fig. 5.** Observed and modelled light curves at 2.25 GHz (crosses, scaled down by 0.3), 2.695 GHz (squares, scaled down by 0.5), 4.75 GHz (plus signs, scaled down by 0.8), 8.3 GHz (asterisks), and 10.55 GHz (diamonds) in July 1993. Source evolution during the event is taken as a smooth increase by about 20% at 8.3 GHz and 10.55 GHz peaking on the 21st of July, in order to agree with the low flux level at 8.3 GHz in the beginning of July 1993. The model light curves for an extreme scattering event harmonise well with the data.

where

$$n_e = 400 \text{ cm}^{-3}, \qquad r_0 = 4 \cdot 10^{11} \text{ cm}. \qquad (1)$$

These numbers are similar to those which Cognard et al. (1993) have deduced for an extreme scattering event in the direction of the pulsar 1937+21, although they used a much harder density profile. Due to the time consuming nature of the numerical analysis of the integral (A13) these numbers are not the result of a fitting process. The fits and the parameters we give therefore illustrate reasonable, not optimal, fits to the data.

We assume this cloud to be a local condensation in the environment of the dark cloud Barnard 30, thereby taking its distance to be 400pc. It has been noted before that extreme scattering events occur preferentially near galactic foreground structure (Fiedler et al. 1994).

The path of the scattering cloud relative to the inner VLBI components of 0528+134 is shown in Fig.6. The outer components C3 and C4 are outside the range of this plot, but still included in the model. The components N and C1 have been shifted according to their apparent motion and the time lag of 0.7 years between our 22 GHz VLBI observation and the event discussed here. The size of all components is taken as independent of frequency and equal to 0.1 mas. A larger size of the components would result in a slight smearing of the modelled light curves. The apparent transverse velocity of the occulter is around 70 km/sec, of which part is due to the motion of the earth around the sun.

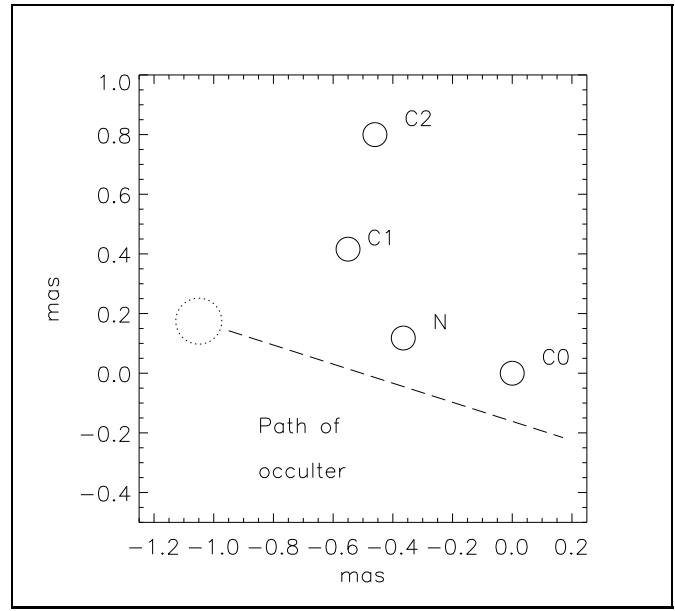

**Fig. 6.** This plot shows the path of the occulter relative to the inner four VLBI components of 0528+134 as used in Fig.5. The components N and C1 have been placed corresponding to their apparent velocity and the time lag of 0.7 years between the 22GHz VLBI observations and the scattering event. The circle's sizes represent the FWHM-sizes of the components, which in our modelling were taken to be independent of frequency. The size of the circle drawn for the occulter represents the size scale $r_0$ and not the overall size of the scattering system.

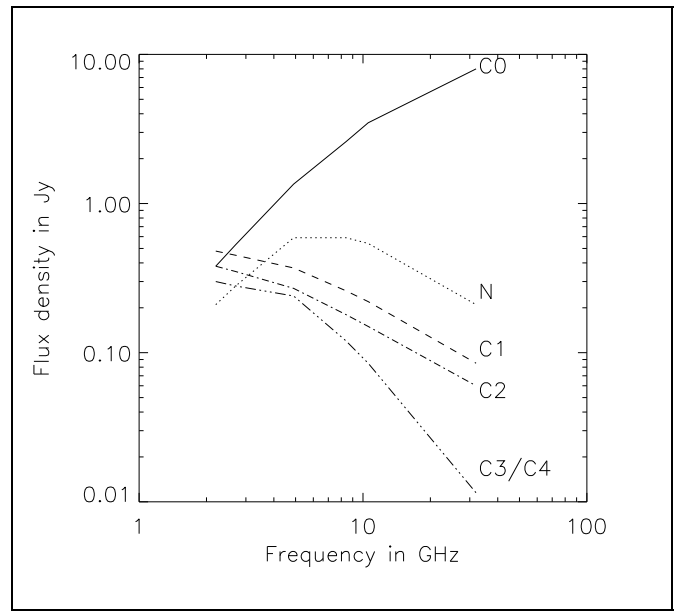

**Fig. 7.** Here we show the spectra of the VLBI components which we used for our modelling. The only significant difference to the observed flux densities of the VLBI components is that the core C0 is assumed to undergo a major outburst in July 1993.

In Fig. 7 we plot the spectra of the VLBI components as taken for our modelling. We assume no significant spectral changes compared to the VLBI results at 22 GHz and 8.4 GHz except for the flaring core component C0. The spectra of all components are taken as stationary for our modelling. For component C2 we are not sure what the spectrum is. While being rather strong at 8.4 GHz in our VLBI data from 1991.23, it does not show up in data at the same frequency taken a month earlier (Z94). The spectrum we use here fits into the time evolution scheme of a cooling electron spectrum in the components. The total flux of the source is chosen to match the observed flux values before (June) and after (August) the event. To do so we have to assume that source evolution increases the 8.4 GHz and 10.55 GHz flux densities smoothly (with a rise time of 10 days) by roughly 20% peaking around the 21st of July following the general trend visible in Fig.2a and 2b.

The observed flux variations in July 1993 can be nicely reproduced by our model. This is especially true for the time lag between the minimum at 10.55 GHz (before 18th of July) and the minima at lower frequencies (after 18th of July). This behaviour is not due to our choice of the density profile in the scattering cloud, but very much related to the VLBI structure of the source. While passing near the line-of-sight the scattering plasma cloud first influences the emission with inverted spectrum from the core component which leads to the early minimum at 10.55 GHz. A few days later the emission from the inner jet components, which have flatter spectra, is scattered yielding later minima at low frequencies. Small fluctuations in the density profile of the scatterer show up preferentially at 2.25 GHz and 2.695 GHz and may cause deviations between the model and the real data, whereas at higher frequencies only the general density profile and the VLBI structure of the source play a rôle.

We regard the observed dips in the light curves of 0528+134 as likely to be explained by an extreme scattering event. It may well be that similar events in the direction of 0528+134 are rather common due to turbulent substructure in Barnard 30.

## 5. Conclusion

In this paper we have reported on multifrequency radio observations between 1992 and 1994 of the $\gamma$-ray quasar 0528+134. In summer 1993 we observed a major outburst at all frequencies higher than a few GHz, which peaked a few months after a strong outburst in high energy $\gamma$-rays and showed the canonical time evolution of a cooling and/or expanding electron distribution. Our data indicate that the outburst in the EGRET range originated very close to the central object of the AGN and that a remnant of this outburst moved further outward in the jet until it became optically thin at radio frequencies after a few months.

We have also presented a new VLBI map at 22 GHz of 0528+134 observed in November 1992. The source structure is similar to previous VLBI maps except for a new component which is not included in a 22 GHz map taken in November 1990 (Z94) but is seen in 43 GHz data from May 1992 (Krichbaum et al. 1995) and at 86 GHz in April 1993 (Standke 1994, PhD Thesis). The apparent transverse velocity of this new component is $\beta_{app}$ ≤20.

A second component can also be identified in the 22 GHz data from 1990 (Z94) and in 8.4 GHz maps. Its apparent transverse velocity is $\beta_{app}$=4.4±1.7 ($H_0$=100, $q_0$=0.5). This result gives evidence for the existence of superluminal motion in the EGRET source, which was expected since Doppler boosting is required to satisfy the compactness limit in terms of medium energy $\gamma$-ray luminosity and variability timescale. Since only few EGRET sources have been looked at with sufficient resolution in VLBI, the high percentage of superluminal sources in the sample of EGRET detected quasars yields strong evidence for the assertion that high energy $\gamma$-ray emission from AGN originates in relativistic jets and is strongly boosted around the direction of the jet.

In July 1993 we observed an unusual strong decrease of the flux density by about 50% at 4.75 GHz and 10.55 GHz and slightly less at 2.695 GHz wavelength. This feature can also be found in the monitoring data at 2.25 GHz and at 8.3 GHz from the Green Bank Interferometer and must be taken as real. An intrinsic origin of the variability behaviour is found to be unlikely on the basis of the source structure, the spectra of its components and the frequency dependence of the dips in the light curves.

Alternatively, we model the event by extreme scattering on a plasma cloud at 400pc distance near Barnard 30. Extreme scattering events are often found in pulsar light curves in which the time-of-arrival variations allow a good distinction between scattering and other effects. We have complemented our light curves by data at 2.25 GHz and at 8.3 GHz from the NRL-GBI program which is designed to investigate extreme scattering events. It is found that the structure of the source is more important for the model light curve than the shape of the scattering plasma cloud is, which in our model was described by a soft, cylinder-symmetric density profile.

While for the stochastic interstellar scintillations the importance of the VLBI structure of the background source for the resulting light curves has already been pointed out by Spangler et al. (1993), in this paper for the first time a model of an extreme scattering event is based on the observed source structure and the spectra of the VLBI components of an AGN. It therefore provides the best available evidence for the existence of extreme scattering events in the light curves of AGN and may together with the statistical information from the NRL-GBI monitoring program help to give insights into the sub-AU scale structure of the interstellar medium in our Galaxy.

## Appendix A: Radiation transport through ionised matter

The phase shift of an electromagnetic wave due to the passage through a cloud of ionised matter at distance $\rho$ to the cloud centre is

$$\Phi(\rho) = \frac{2\pi r_e c}{\omega} \int_{-\infty}^{\infty} dl\, n_e = \frac{\Phi_0}{r_0 + \rho} \quad , \tag{A1}$$

where we have assumed an axisymmetric density profile

$$\int_{-\infty}^{\infty} dl\, n_e = \frac{n_e\, r_0^2}{r_0 + \rho} \quad .$$

In reality such a profile can not be regarded as extending to infinity, but will be immersed in small scale density fluctuations. The density profile (A1) is still a good approximation if it dominates the true profile including fluctuations only at radii smaller than $\rho_d$ where

$$\rho_d \gg \left(2\pi n_e \lambda^2 z r_e r_0^2\right)^{1/3} \simeq 3 \cdot 10^{12} \left(\frac{\lambda}{11\,\mathrm{cm}}\right)^{2/3} \quad \mathrm{cm} \quad (A2)$$

for our parameters. Our description will of course also not work for $x \geq \rho_d$, i.e. when the line-of-sight to the AGN traverses regions far from the scatterer, which are dominated by small scale density fluctuations, around a month before or after the event. The mass contained by the scatterer is

$$M_{sc} \approx \frac{7 \cdot 10^{14}}{f_{ion}} \left[\frac{\rho_d}{r_0} - \ln\left(1 + \frac{\rho_d}{r_0}\right)\right] \quad \mathrm{grams} \; . \quad (A3)$$

Here $f_{ion}$ denotes the degree of ionisation.

With the wave propagator in the Fresnel regime (Born and Wolf 1975) we obtain for the amplitude function (relaxing the normalisation)

$$A(x) = \frac{\sqrt{\lambda z}}{\pi} \int_0^\pi d\phi \int_0^\infty ds$$
$$\times \exp\left[i\left(s - \sqrt{\frac{s}{s_c}}\cos\phi - \frac{\Phi_1}{r_f + \sqrt{s}}\right)\right] \; , \quad (A4)$$

where $x$ is the distance at the cloud between the observer's line-of-sight to the AGN and the symmetry axis through the AGN and the cloud, $z$ is the distance between observer and the scatterer, $s = \frac{\pi \rho^2}{\lambda z}$, $\Phi_1 = \sqrt{\frac{\pi}{\lambda z}}\Phi_0$, $r_f = r_0\sqrt{\frac{\pi}{\lambda z}}$ and $s_c = \frac{\lambda z}{4\pi x^2}$.

We use the method of stationary phase for the s-integral (e.g. Wong 1989).

$$f(s) = s - \sqrt{\frac{s}{s_c}}\cos\phi - \frac{\Phi_1}{r_f + \sqrt{s}} \quad (A5a)$$

$$f'(s) = 1 - \frac{\cos\phi}{2\sqrt{s_c s}} + \frac{\Phi_1}{2\sqrt{s}(r_f + \sqrt{s})^2} \quad (A5b)$$

$$f''(s) = \frac{\cos\phi}{4\sqrt{s_c}} s^{-3/2} - \frac{\Phi_1}{4s^{3/2}(r_f + \sqrt{s})^2} - \frac{\Phi_1}{2s(r_f + \sqrt{s})^3}$$
$$(A5c)$$

The condition $f'(s) = 0$ gives a polynomial of third order in $y = \sqrt{s}$.

$$y^3 + \left(2r_f - \frac{\cos\phi}{2\sqrt{s_c}}\right)y^2 + \left(r_f^2 - \frac{r_f \cos\phi}{\sqrt{s_c}}\right)y$$
$$+ \frac{\Phi_1}{2} - \frac{r_f^2 \cos\phi}{2\sqrt{s_c}} = 0 \; . \quad (A6)$$

With the variable $x = y + \frac{2r_f}{3} - \frac{\cos\phi}{6\sqrt{s_c}}$ we obtain the reduced equation

$$x^3 + px + q = 0 \quad (A7)$$

where

$$p = -\frac{r_f^2}{3}(1+\epsilon)^2 \; , \qquad \epsilon = \frac{x\cos\phi}{r_0} \quad (A8)$$

$$q = -\frac{2}{27}r_f^3\left[1 + 3\epsilon + 3\epsilon^2 + \epsilon^3 - \eta\right] \; , \qquad \eta = \frac{27\Phi_0}{4r_0 r_f^2} \; . \quad (A9)$$

For the discriminant $D = (p/3)^3 + (q/2)^2$ we derive

$$D = 3^{-6} r_f^6 \left[\eta^2 - 2\eta(1+\epsilon)^3\right] \; . \quad (A10)$$

Depending on the sign of $D$ we have one or three real zeros $y_0$ of the above equation. Interesting for us are only the positive ones. The $\Phi$-integral is then carried out only for those $\Phi$ for which a positive zero exists.

With the reduced zero $y_0^* = \frac{y_0}{r_f}$ we get for the coefficients of the expansion

$$f(s_0 = y_0^2) = r_f^2 \left(y_0^{*2} - 2\epsilon y_0^* - \frac{4}{27}\frac{\eta}{1+y_0^*}\right) \quad (A11)$$

and

$$f''(s_0) = r_f^{-2}\left[\frac{\epsilon}{2y_0^{*3}} - \frac{\eta(1+3y_0^*)}{27\,y_0^{*3}(1+y_0^*)^3}\right]$$
$$= r_f^{-2} g(y_0^*) \; . \quad (A12)$$

For the amplitude function we then get

$$A(x) \simeq \sqrt{2}r_0 \int_0^\pi d\Phi$$
$$\times \sum_{y_0^*} \frac{1}{\sqrt{\pm g(\Phi)}} \exp\left(i\left[f(\Phi) \pm \frac{\pi}{4}\right]\right) \quad (A13)$$

which can be solved numerically. Here the sign is determined by $\mathrm{sgn}\,g(\Phi)$ in the sense that the argument of the square root must be positive.

## Appendix B: Luminosity limits in the extreme relativistic range

There are basically two luminosity limits for AGN, the Elliot-Shapiro relation and the compactness limit. The former is the Eddington limit written in terms of the variability timescale $\tau$ as light travel time for the Schwarzschild radius of the source. In the relativistic limit the Thomson cross section is no longer valid and one has to use the Klein-Nishina cross section. Then

$$L \leq \frac{8\tau m_p c^4}{\sigma_T}\frac{\epsilon}{\ln(2\epsilon)} \; , \qquad \epsilon = \frac{E_\gamma}{m_e c^2} \quad (B1)$$

This limit is not violated during the outburst of 0528+134 in March 1993, for which an integrated photon flux of $I(E > 100\,\mathrm{MeV}) \simeq 3 \cdot 10^{-6}\,\mathrm{cm^{-2}sec^{-1}}$ is reported (Michelson et al. 1994). The spectral index was $s = 2.25 \pm 0.11$ and there is weak evidence for short-time variability. This limit was also not violated in May 1991 when the source was detected for the first time. At that time the integrated photon flux was a factor 2.5 less, the spectral index was about 2.6, and clear evidence was found for variability on timescales of less than a day in the source frame (Hunter et al. 1993).

Now we discuss the compactness limit in the relativistic range. The optical depth of photons with energy $\epsilon_1$ (here all energies are in units $m_e c^2$) due to pair production by photon-photon interactions in a spherical emission region of radius $R$ and differential photon luminosity $L(\epsilon)$ is

$$l(\epsilon_1) \simeq R \oint d\Omega (1-\mu) \int d\epsilon' \, n'(\epsilon', \Omega) \, \sigma(\epsilon') \qquad (B2)$$

where

$$n'(\epsilon', \Omega) = \frac{3}{16\pi^2 R^2 c} \frac{4\epsilon'}{\epsilon_1(1-\mu)} L\left(\frac{2\epsilon'^2}{\epsilon_1(1-\mu)}\right) \qquad (B3)$$

is the differential photon density in the CMS of the photon-photon system and

$$\sigma(\epsilon') = \frac{\pi r_0^2}{2}(1-\beta^2)\left[(3-\beta^4)\ln\frac{1+\beta}{1-\beta} - 2\beta(2-\beta^2)\right] \qquad (B4)$$

with

$$\beta = \sqrt{1 - \epsilon'^{-2}}$$

is the total cross section for pair production in the CMS (Jauch and Rohrlich 1976).

In case of a spectrum

$$L = L_0 E_{min} \epsilon^{-2}, \quad \epsilon \geq E_{min} \qquad (B5)$$

we get

$$l(\epsilon_1) = \frac{3 L_0 E_{min} \epsilon_1}{8\pi \tau c^2} \int_{-1}^{1} d\mu \, (1-\mu)^2 \int_{\epsilon_{min}} d\epsilon' \, \frac{\sigma(\epsilon')}{\epsilon'^3} \qquad (B6)$$

with

$$\epsilon_{min} = \max\left(1.0, \sqrt{0.5 E_{min} \epsilon_1 (1-\mu)}\right)$$

and $\tau$ as the minimal variability timescale. In a $q_0 = 0.5$ cosmology we can write the luminosity in terms of the observed differential photon spectrum $I(E) = I_0 E'_{min} E^{-2}$

$$L_0 E_{min} = \frac{16\pi c^2 (\zeta^{0.5}-1)^2 I_0 E_{min}}{H_0^2} \qquad (B7)$$

where

$$\zeta = 1 + z, \quad E_{min} = \zeta E'_{min}$$

Then

$$l(\epsilon_1) = \frac{6(\zeta^{0.5}-1)^2 I_0 E_{min} \epsilon_1}{\tau H_0^2}$$

$$\times \int_{-1}^{1} d\mu \, (1-\mu)^2 \int_{\epsilon_{min}} d\epsilon' \, \frac{\sigma(\epsilon')}{\epsilon'^3} \qquad (B8)$$

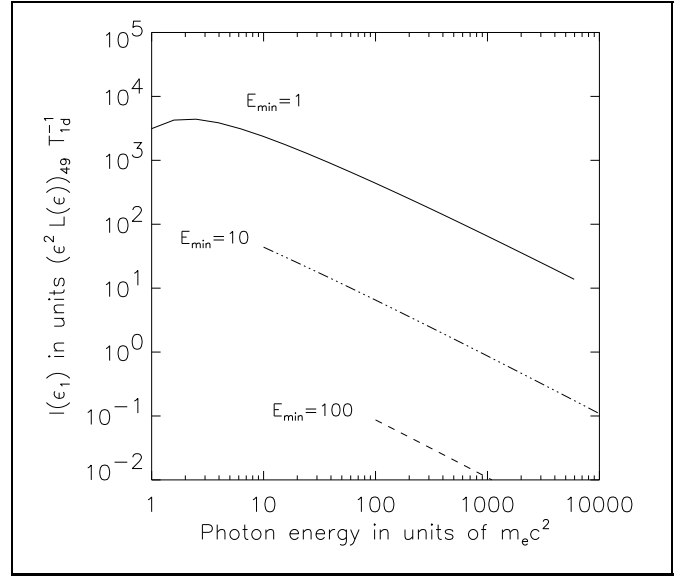

**Fig. 8.** Compactness of an AGN in dependence of the low energy cut-off of its gamma ray spectrum. Here $E_{min}$ is in units of $m_e c^2$. The compactness scales linearly with the $EF(E)$-luminosity in units of $10^{49}$ ergs/sec and inversely with the variability timescale in the frame of the AGN in units of days. For 0528+134 the scale factor is around unity when $H_0 = 100$ km s$^{-1}$ Mpc$^{-1}$, $q_0 = 0.5$ is assumed.

In Fig.8 we have plotted the integral (B8) for different low-energy cut-offs in the emission spectrum of 0528+134. There are 2.5$\sigma$ detections at 3-30 MeV from COMPTEL for the viewing periods in May 1991 and in March 1993 (Collmar et al. 1994 and private communication), which indicate a continuation of the EGRET spectra down to a few MeV and flatter spectra at lower energies. In May 1993 0528+134 was not detected by COMPTEL. One should note that when adding all data 0528+134 is detected by COMPTEL with a significance of more than 5$\sigma$, so it is clearly there, although for single viewing periods there are three periods with ~2.5$\sigma$ detections and a few with less than 1$\sigma$. It is thus reasonably probable that the three ~2.5$\sigma$ detections are real, although the significances for the single observations are very small. The nonlinear dependence of the significance of a source on the square root of the observing time is due to COMPTEL-specific background problems.

One should note that on the basis of EGRET data alone one would not be able to see the signature of such an attenuation since in an optically thick situation the effective optical depth depends on the attenuated spectrum of photons. The cut-off energy can be roughly determined from Fig.8. It is that value of $E_{min}$ for which the compactness at photon energy $E_{min}$ is $l = 1$. In our case this is $E_{min,c} \simeq 50$ which corresponds to $E_c \simeq 8$ MeV in the observer frame.

Since the variation of the photon flux in the COMPTEL range follows the general trend seen by EGRET, it is natural to assume that the $\gamma$-rays in both energy regime originate in the same region. From Fig.8 we can deduce that the photons with energies less than 10 MeV (observer frame) should have been strongly attenuated if the emitting regions were stationary, which is in contradiction to their detection by COMPTEL.

Therefore, Doppler boosting of the **medium** energy γ-ray emission of 0528+134 is required.

*Acknowledgements.* MP acknowledges financial support from the DARA Verbundforschungsprogramm (50 OR 9301 1) during his stay in Bonn. We thank W. Collmar for many discussions on the COMPTEL data and for the possibility to use his newest results. We like to thank the staff at the Effelsberg 100-m telescope, the IRAM 30-m telescope and R. Kothes for assistance with the observations. RS gratefully acknowledges partial support by the DARA (50 OR 9406 3) of his Compton observatory guest investigator programs. We further like to thank A. Witzel and C.J. Schalinski for useful discussions and comments, and A.M. Gontier for her help in the reduction of the geodetic VLBI data. T.P.K. acknowledges the support of the German BMFT Verbundforschung. S.B. was supported in part by the DFG. K.S. thanks for the financial support of the Geodätische Institut der Universität Bonn. We also like to thank Alan Marscher and Yun Fei Zhang for the provision of their VLBI maps prior to publication.